\documentclass[rmp,twocolumn]{revtex4}

\usepackage{graphicx,amsmath,amssymb,txfonts,dcolumn}

\begin{document}

\title{An integrated model of traffic, geography and economy in the
  Internet}

\author{Petter Holme}
\affiliation{School of Computer Science and Commuication, Royal
  Institute of Technology, 100 44 Stockholm, Sweden}
\affiliation{Department of Computer Science, University of New Mexico,
  Albuquerque, NM 87131, U.S.A.}

\author{Josh Karlin}
\affiliation{Department of Computer Science, University of New Mexico,
  Albuquerque, NM 87131, U.S.A.}

\author{Stephanie Forrest}
\affiliation{Department of Computer Science, University of New Mexico,
  Albuquerque, NM 87131, U.S.A.}
\affiliation{Santa Fe Institute, 1399 Hyde Park Road, Santa Fe, NM
  87501, U.S.A.}

\begin{abstract}
Modeling Internet growth is important both for understanding the
current network and to predict and improve its future. To date,
Internet models have typically attempted to explain a subset of the
following characteristics: network structure, traffic flow, geography,
and economy. In this paper we present a discrete, agent-based model,
that integrates all of them. We show that the model generates networks
with topologies, dynamics, and (more speculatively)
spatial distributions that are similar to the Internet.
\end{abstract}

\maketitle

\section{Introduction}

As one of the most complex human constructions, the Internet
is a challenging system to model. Dynamic processes of different
time-scales operate simultaneously---from slow processes, like the
development of new hardware, to the transport of data, which occurs at
the speed of light.

These phenomena are to some extent interdependent.  Traffic provides
income to the service providers which is then invested in
infrastructure, which can lead to changes in traffic patterns.  This
paper describes an agent-based model (ABM) that attempts to reproduce
large-scale features of the Autonomous System (AS) level of the
Internet by modeling localized and well-understood network
interactions.  The ASes of the Internet lend themselves naturally to
discrete ABM models~\cite{Bonabeau02a}.  Each AS is an economic agent,
comprised of a spatially discrete network.  Over time, ASes create new
links to other ASes, upgrade their carrying capacity, and compete for
customer traffic.  The agents in the model described here, behave
similarly, although we have simplified as much as possible.
Specifically, the model is designed to be both simple and general
enough to simulate any spatially extended communication network built by
subnetworks of economically driven agents.

In previous work, Chang \emph{et al.}\ showed that incorporating
economics and geography into the Highly-Optimized Tolerance
(HOT)~\cite{carlson:hot} model increases the model's
accuracy~\cite{chang:superhot}.  A related ABM model of the AS graph
produces degree distributions similar to empirical
observations~\cite{peer:chang}.  Bar~\textit{et al.}\ proposed a
similar model~\cite{bar:inet}, that incorporates another aspect of the
real Internet---that the agents are spatially extended objects.  Our
model is similar in scope to this earlier work but differs in the
details, most importantly by adding explicit economics (cost) to the
model. Other differences include accounting for population density,
simplifying the treatment of traffic flow, and not assuming a HOT
framework.  The previous work in this area, like much research on
network models, focuses almost exclusively on degree distributions of
the graphs.  In this paper, we compare our results to Internet data
using several topological measures~\cite{our:rad}, including degree
distributions, as well as geography and traffic dynamics.

The remainder of the paper is organized as follows. First, we describe
and motivate the model. Then, we characterize the time evolution,
network topology, correlation between network structure and traffic
flow, packet routing statistics, and geographical aspects of the
networks produced by the model. Where possible, we compare the
properties of these synthetic networks to observed data from the
Internet.

\section{AS simulation model}

We begin with the fundamental unit responsible for
network growth, an agent with economic interests~\cite{gao:relation}.
These agents manage traffic over a geographically extended network
(which we refer to as a \textit{sub-network} to distinguish it from the network
of ASes) and profit from the traffic that flows through their network.

We compare the agents to the ASes that comprise the Internet.  This is
not an exact mapping---some of the Internet Service Providers (ISPs)
have many AS numbers (e.g., AT\&T), while other ASes are shared by
several organizations. We make the common simplifying assumption that
once an agent is introduced, it does not merge with another agent or
go bankrupt~\cite{vesp:inet,claffy:aseco,peer:chang}. This is
partially justified by the fact that the Internet, from its inception,
has grown monotonically, and we seek to capture this dynamic in our
model Most other models of the AS graph enforce strict
growth~\cite{vesp:inet} as well and are, as ours, justified by their
\textit{a posteriori} ability to reproduce measured features.

We assume a network user population distributed over a two-dimensional
area. Traffic is simulated by a packet-exchange model, where a
packet's source and destination are generated with a probability that
is a function of the population profile. The model is initialized with
one agent comprised of a network (a sub-network in our terminology)
that spans one grid location (referred to as a {\em pixel} of the
landscape. As time progresses, the agent may extend its subnetwork to
other pixels, so that the sub-networks reach a larger fraction of the
population. This creates more traffic, which generates profit, which
is then reinvested into further network expansion.  Through positive feedback, the network
grows until it covers the entire population. In this section we
describe the assumptions and most of the details of the model; the
source code is publicly available from
www.csc.kth.se/$\sim$pholme/asim/.

\begin{figure}
  \centering\resizebox*{0.8 \linewidth}{!}{\includegraphics{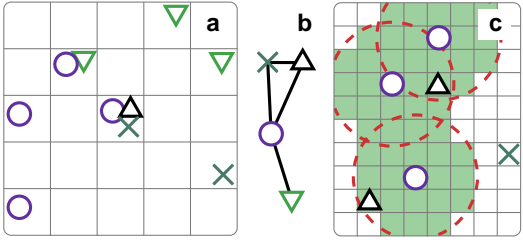}}
  \caption{Illustration of the network growth algorithm. (a) shows the
    locations of four agents on the geographic grid. These are assumed
    to be connected by a physical network administrated by the agent,
    but is not explicit in the model. (b) is an example graph
    resulting from (a). That two agents are present in the same pixel
    is a necessary, but not sufficient condition for a link to form
    between the agents. (c) illustrates the area that each
    hypothetical agent can afford to expand to (the shaded region).}
  \label{fig:nwk}
\end{figure}

An agent $i$ is associated with a set of locations $\Lambda_i$
(representing sources or end-points of traffic, and peering points), a
capacity $K_i$ (limiting the rate of packets that can pass through the
agent), a packet-queue $Q_i$, and a set of neighbor agents
$\Gamma_i$. A necessary, but not sufficient, condition for two agents
to be connected is that their locations overlap at, at least, one
pixel. The locations exist on an $L_x\times L_y$ square grid. A pixel
of the grid is characterized by its population $p(x,y)$ and the set of
agents with a presence there $\mathcal{A}(x,y)$. The total number of
agents is denoted by $n$, and the number of links between agents by
$m$. These quantities, except $L_x$ and $L_y$, depend on the
simulation time. The outer loop of the model then iterates over the
following steps:
\begin{enumerate}

\item \label{step:nwk} \textit{Network growth.} The number of agents
  is increased. Existing agents expand geographically, and their
  capacities are adjusted.

\item \label{step:traffic} \textit{Network traffic.} Packets are
  created, propagated toward their targets, and delivered. This
  process is repeated $N_{\mathrm{traffic}}$ times before the next
  network-growth step.

\end{enumerate}
We measure simulation time $\tau$ as the number of times
Step~\ref{step:nwk} is executed (the time unit between packet
movements is $1/N_{\mathrm{traffic}}$). In the remainder of this
section we describe the growth and traffic steps in greater detail.
 
\subsection{Network growth}

The income of an agent, during a time step, is proportional to the
traffic propagated by the agent during the period. This is a
simplification---in a more detailed simulation one could let the
income depend both on the amount of traffic, and the prices for
forwarding the packets set by business agreements. Assume an agent $i$
has a budget $B_i$ that it tries to invest so that it can increase its
traffic, and thus its profit. Since there is a possibility of
congestion in the model, agent $i$ tries first to remove bottlenecks
by increasing its capacity $K_i$ (the number of packets that the agent
can transit during one time step). When the capacity is sufficient,
the agent spends the rest its budget on increasing its traffic by
expanding geographically. There are three prices associated with
network growth. First, the capacity price
$C_{\mathrm{capacity}}$---the price of increasing $K_i$ one unit. For
simplicity we let $C_{\mathrm{capacity}}$ be independent of the size
of the agent's subnetwork. Second, the wire price
$C_{\mathrm{wire}}$. This is the price per pixel between a new
location and the agent's closest existing location. Last, the cost
$C_{\mathrm{connect}}$ to connect two agents with locations at the
same pixel.

It has been observed that the average degree (number of neighbors of
an AS) in the AS graph is relatively constant over
time~\cite{vesp:inet,daub:as}. We take this as a constraint in our
model and let the desired average agent degree $k_D$ be a control
parameter. We also assume that each agent tries to spend all of its
budget, but not more than that, whenever it is updated.

The network growth step iterates over the following steps:

\begin{enumerate}

\item \label{step:add_as} \textit{Increase of the number of agents.}
  As long as the network is too dense (i.e.\ if $2m>k_Dn$), new agents
  are added. New agents are situated in the pixel $(x,y)$ that has the
  highest available population $p(x,y)/(A(x,y)+1)$ where $A(x,y)$ is
  the cardinality of $\mathcal{A}(x,y)$ and $A(x,y)\geq 1$.  The
  budget and capacity of the new agents are initialized to $B_\mathrm{init}$ and
  $K_\mathrm{init}$ respectively.

  If the network is small, $n<k_D+1$, it is not dense enough for new
  agents to be added in step~\ref{step:add_as}.  Thus, we do not apply
  this condition when $n$ is less than a threshold $n_0$ and call the
  time when $n=n_0$ is reached $t_0$.

\item \label{step:capacity} \textit{Capacity increase.} Each agent
  synchronously increases its subnetwork's capacity based upon traffic
  from the last time step (but not more than the agent can afford).
  Agent $i$ invests the minimum of ($B_i$, $C_{\mathrm{capacity}}
  \Delta T_i,0$, 0) to increase capacity ($\Delta T_i$ is the change in
  traffic propagated by $i$ since the last update).

\item \label{step:link} \textit{Link addition.} While $2m\leq nk_D$
  (which usually means $k_D-1$ times), choose two agents randomly that are not
  already connected and share a common pixel. If the budgets of
  both agents are larger than $C_{\mathrm{connect}}$, then connect
  them.

\item \label{step:extend} \textit{Spatial extension.} Let the agents
  with remaining budget to spend extend their networks. Iterate
  through all agents $i$ and add a location at the pixel, not in
  $\Lambda_i$, that has the highest available population
  $p(x,y)/(L(x,y)+1)$, and is not further than
  $(B_i-C_{\mathrm{connect}}) / C_{\mathrm{wire}}$ from a location in
  $\Lambda_i$ (i.e., not further from $i$ than $i$ can afford). (See
  Figure~\ref{fig:nwk}(b)).  An alternative location selector might
  select the point which has the lowest cost per unit of population.
  Unfortunately, such an algorithm is computationally prohibitive for
  modeled networks of the Internet's scale.
\end{enumerate}

\begin{figure}[t]
  \centering\resizebox*{0.8 \linewidth}{!}{\includegraphics{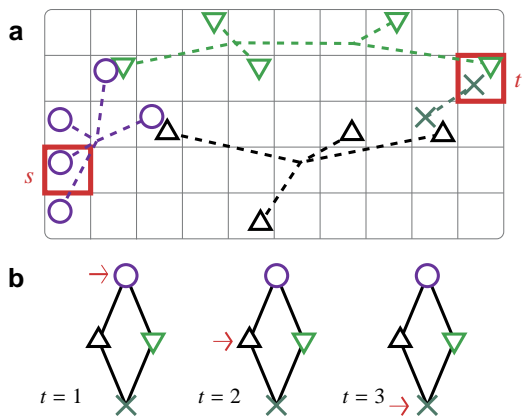}}
  \caption{Illustration of traffic simulation. (a) A packet is created
    with source pixel $s$ and target pixel $t$ with probability
    proportional to the product of populations at $s$ and $t$. One of
    the agents at the target pixel is randomly chosen as the target
    agent. The propagation of the packet is shown in the graph.  Each
    agent $i$ is associated with a queue $Q_i$ and a capacity
    $K_i$. When a packet reaches an agent, it is appended to
    $Q_i$. $K_i$ packets in the queue are relayed to neighboring
    agents and $i$'s budget is credited one unit. The arrows in (b)
    symbolize the packet's route from source to destination agent. The
    package is routed to a neighboring agent $j$ with probability
    $\exp((d(i,t)-d(j,t))/\lambda$ (where $t$ is the packet's target,
    $d(\:\cdot\:,\:\cdot\:)$ gives the graph distance, and $\lambda$
    is a parameter).
  }
  \label{fig:dyn}
\end{figure}

The cost of each agent modification mentioned above is immediately
deducted from the budget of the agent.

\subsection{Network traffic}

We model traffic with a discrete, packet-exchange
model~\cite{holme:traffic,echen:cong}. The packets are generated with
specific source and target pixels, but the routing takes place on
the network of agents. We neglect intradomain routing between the
agent's locations, assuming the time it takes for a packet to pass
through an agent is independent of the specific locations it
visits. The dynamics are defined as follows:

\begin{enumerate}

\item \label{step:create} \textit{Packet generation.} We assume that
  most traffic originates from direct communication between
  individuals and does not depend on the distance between them.  So,
  for each pair of points $[(x,y),(x',y')]$ on the grid, we create a
  packet with source $(x,y)$ and destination $(x',y')$ with
  probability $P_{\mathrm{pkg}}\,  p(x,y)\, p(x',y')$. Then one agent,
  selected at random from the agents with a location at the pixel, is
  made the source node for the packet. The destination agent is
  randomly chosen from the agents at the destination pixel. Finally,
  one unit of credit is added to the sender's budget.
  
\item \label{step:propagate} \textit{Packet propagation.}  Each agent
  $i$ propagates the first $K_i$ packets from its queue (of length
  $l_i$) each time step and receives one unit credit for each
  propagated packet. A packet can propagate only one hop (inter-AS
  transmission) per time step. A packet at agent $i$ is propagated to
  a neighbor $j$ with probability $\exp(\lambda(d(i,t)-d(j,t))$ (where
  $t$ is the recipient AS, $d(\:\cdot\:,\:\cdot\:)$ is the graph
  distance, and $\lambda$ is a parameter controlling the deviation
  from shortest-path routing~\cite{sood:rw} observed in
  Ref.~\cite{gao:inflate}).
  
\item \label{step:delete} \textit{Packet delivery.} For all agents,
  delete all packets that have reached their target.
  
\end{enumerate}

The assumption, in step~\ref{step:create}, that the probability that
two agents communicate is independent of their spatial separation is
in line with the (somewhat debated) ``death of distance'' in the
Internet age~\cite{cairncross}. We also tested communication rates that
decay with the square of the distance, as observed in conventional
trade firms~\cite{isard}, with qualitatively the same results.

Business agreements between ASes are an important factor in the Border
Gateway Protocol (BGP)~\cite{rfc1771} (the Internet's largest scale
routing protocol). Next hops are often selected by cost, rather than
path length.  We do not explicitly include inter-AS contractual
agreements, but our probabilistic propagation
method~\ref{step:propagate} has a similar effect on average path
length~\cite{gao:inflate}.

\section{Numerical simulations}

\subsection{Parameter values}

Before presenting the simulation results, we describe the experimental
design, and choice of parameters. First, we specify a population
profile $p(x,y)$. We primarily model population distributions, but we
also model specific geographic populations (e.g. U.S.A. census data).
To simplify the generation of population distributions, we neglect
spatial correlations and simply model the frequency of population
densities. This frequency has two important features: it is skewed
(pixels with low population densities are more frequent than highly
populated pixels) and fat-tailed (there are pixels with a population
density many orders of magnitude larger than the average). One
probability distribution with such features is the power-law
distribution $\mathrm{Prob}\: p \sim p^{-\chi}$.  To reduce the
fluctuations between different realizations of $\{p(x,y)\}$, and
prevent unrealistically high populations within a pixel, we sample the
power-law distribution in the bounded interval $[1,(L_x
  L_y)^{1/(1-\chi)}]$~\cite{chn:perc} with $\chi=3$. Our results do
not depend strongly on the distribution $p(x,y)$.  We obtain
qualitatively similar result with normally distributed $p$-values and
real population-density maps (data not shown).

\begin{table}[t]
  \centering
  \begin{tabular}{|c|c|c|}
    \hline
    Parameter & Interpretation & Value \\
    \hline
    $L_x=L_y$ & Number of pixels in the x (and y) direction & 50 \\
    $N_{\mathrm{traffic}}$ & Number of packets sent per simulation step & $1\times 10^4$ \\
    $P_{\mathrm{pkg}}$ & Constant to determine packet source and dest. & $0.001$ \\
    $n_0$ & Agent growth threshold & $35$ \\
    $K_\mathrm{init}$ & Initial capacity of an agent & $5$ \\
    $C_{\mathrm{wire}}$ & Price per pixel for new wire & $500$ \\
    $B_\mathrm{init}$ & Initial budget for a new agent & $3\times 10^5$ \\
    $\lambda$ & Parameter in exponential distribution & $75$ \\
    \hline
  \end{tabular}
  \caption{Default parameters values for simulation experiments.}
  \label{tab:params}
\end{table}

In multiparameter, agent-based models, such as ours, a systematic
investigation of the full parameter space is infeasible. Parameters
are, if possible, obtained from real systems.  We set the desired
degree $k_D=5.52$ as observed in Ref.~\cite{our:rad}. Unless otherwise
stated, the desired size of the network is $n_D=16{,}000$, which is
the same order of magnitude as the real AS graph. Other parameters are
balanced to keep runtime low (less than one day) while still engaging
all aspects of the algorithm.  This means, for example, that between
every network update, a significant number of packets are routed
through even the smallest agents, and enough packages to cause
congestion pass through larger agents. Unless otherwise stated, we use
the parameter set given in Table~\ref{tab:params}.  Many of the
results we show are from a single run, we have confirmed that the
results are representative by comparing them with 20 other runs.

\begin{figure}
  \centering\resizebox*{0.95 \linewidth}{!}{\includegraphics{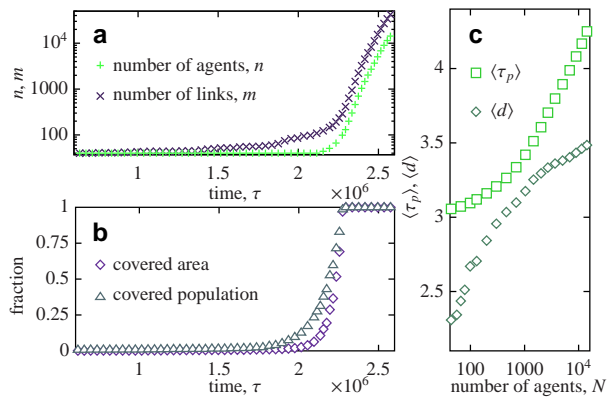}}
  \caption{Time evolution of an example run. In panel (a) the number
    of agents and the number of inter-agent links as a function of
    simulation time. In (b) the fraction of the landscape with network
    coverage, and the fraction of the population reached by the
    network, is plotted against time. Panel (c) shows the average
    travel time $\langle\tau_p\rangle$ for packets and the average distance (number of
    inter-agent hops) in the network $\langle d\rangle$, as functions of the number of
    agents.  }
  \label{fig:tme}
\end{figure}

\subsection{Network Growth}
\label{subsec:growth_results}
We begin by studying the growth of the network over time. In
Fig.~\ref{fig:tme}(a) we plot the number of agents and links as a
function of simulation time for one representative run. At
$\tau=\tau_0\sim 4\times 10^5$ the graph is sparser than
$k_D$. Initially, the agents spend the budget they accumulate on new
links (and increasing capacity). Around $\tau\sim 1.5\times 10^6$, the
budget of the wealthier agents is sufficient to invest in wires to new
locations (see Fig.~\ref{fig:tme}(b)). This creates new traffic, which
causes positive feedback accelerating the traffic flow, coverage,
budget, and also more congestion. Around $\tau\sim 1.9\times 10^6$,
$n(\tau)$ and $m(\tau)$ change from exponential to sub-exponential
growth. As we see below, this is also the time when a significant
level of congestion appears in the system. At about the same time, the
entire population is serviced by the network. With the current model,
the network would grow indefinitely but with decreasing returns for
the agents. Alternatively one could introduce maintenance costs
proportional to network size, in which case the network would reach a
steady state where the budgets of the agents are balanced and no
further investments can be made.  For $\tau\gtrsim 1.9\times 10^6$ the
increase of $n(\tau)$ is slower than exponential. This is explained by
the increasing level of congestion in the system. In
Fig.~\ref{fig:tme}(c) we plot the average time $\langle\tau_p\rangle$
for a packet to travel from source to
destination. $\langle\tau_p\rangle$ is bounded from below by the
average distance (number of links in the shortest path, averaged over
pairs of nodes) $\langle d\rangle$. The two curves diverge, i.e.\ a
significant level of congestion appears, around $N=1000$. The growth
of $n(\tau)$ and $m(\tau)$ slows down at the same point. We conclude
that growth slowdown comes from a congestion-driven negative feedback.
The most striking feature of network
growth over time is the transition from a small network, almost
constant in size, to a rapidly increasing system (around $\tau \sim
1.8\times 10^6$). This effect is typical for technologies emerging
from the interactions of a large number of agents---they need a
\emph{critical mass} of users to reach a significant fraction of the
total population. One can argue that the Internet reached this
critical mass in the early 1980's when it started to span the
globe. Another important point in the Internet's history was the advent of
the World Wide Web (WWW) in the early 1990's, and with it commercial
applications and access to the general public. Our model does
not include applications, such as the WWW, that undeniably affect network growth. Such effects could be included by adopting a different traffic model,
but for this paper we aim at simplicity and generality. In the
Internet the growth of the number of ASes is slower than the
exponential increase of agents predicted by the
model (bgp.potaroo.net/cidr/; read January 7, 2008). One reason
for the faster growth is that we do not assume that maintenance costs
are proportional to income---if such costs grow super-linearly,
negative feedback could dampen growth. Other external factors, such as
the fact that AS numbers are allocated and assigned by a central
authority (Internet Assigned Numbers Authority, www.iana.org), might
also influence the actual rate of growth experienced by the Internet.

\begin{figure}
  \centering\resizebox*{0.95 \linewidth}{!}{\includegraphics{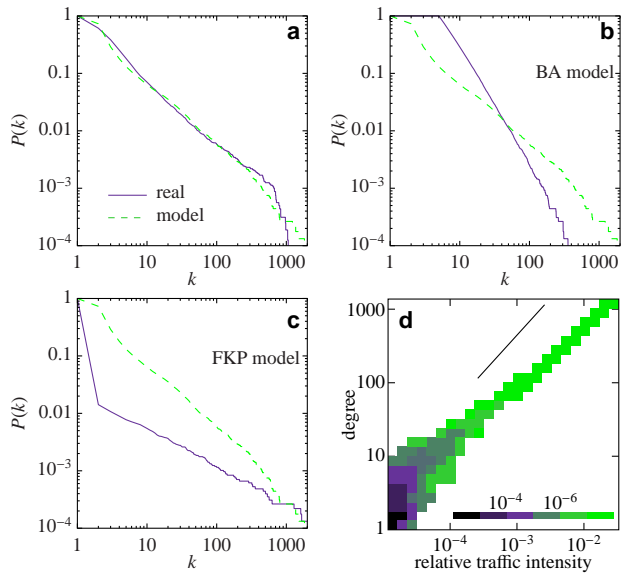}}
  \caption{The degree distribution (cumulative mass function) of a
    real AS-graph (AS06) together with degree distribution of a
    network generated with the model (a), the BA (b) and the FKP
    models (c). Panel (d) is a density plot that illustrates the
    correlation between traffic and degree in our model runs.}
  \label{fig:deg}
\end{figure}

\subsection{Degree distribution}

One of the most conspicuous network structures of AS-graphs is its
skewed degree distribution (first observed in Ref.~\cite{f3}),
compatible with a power-law functional form~\cite{aaron:power_law}. In
Fig.~\ref{fig:deg}(a) we compare the cumulative degree distribution of
our model with that of the Internet's. We use the model network from
the example run described earlier (taking data from the simulation
when $N=16{,}000$), and the ``AS06'' network of Ref.~\cite{our:rad}
(an AS-graph constructed from www.routeviews.org and www.ripe.net,
with $N=22{,}688$). The match between the model and the real networks
is striking.  Preliminary studies indicate that the slope of the curve
is largely insensitive to changes in parameter values. We compare this
result with a generic network model that produces power-law degree
distributions (the Barab\'{a}si--Albert (BA) model~\cite{ba:model})
and a simple, geographic model of the AS-graph designed by Fabrikant,
Koutsoupias, and Papadimitriou (FKP)~\cite{fkp:model}.

The BA model is a growth model in which one node (and $m$ links to
attach it with the rest of the network) is added every time
step. \textit{Preferential attachment} is used to determine the
endpoints of the new links---the probability of attaching to a node of
degree $k$ is proportional to $k$.

The FKP model is also a simple growth-model.  Each time step, one
node, and a link attached to it, is added to the graph. A new node $i$
is assigned random coordinates in the unit square and attached to the
old node $j$ that minimizes $d_0(j)+\alpha
|\mathbf{r}_i-\mathbf{r}_j|$ (where $d_0(j)$ is the graph distance
between $j$ and the node added first,
$|\mathbf{r}_i-\mathbf{r}_j|$ is the Euclidean distance between $i$
and $j$, and $\alpha$ is a parameter setting the cost-balance between
making new physical connections or using the existing network).

In Figs.~\ref{fig:deg}(b) and (c) we plot the cumulative mass function
of degree for one BA and one FKP network. The model parameter values
were chosen to give networks as close as possible to the real AS-graph
($m=5$ for the BA model, $\alpha=4$ for the FKP model, and $N=22{,}688$
for both). The slope of the BA model is steeper than the real
network, and the curve for the FKP-model is flatter than the real
data.  To compare the goodness-of-fit, since the curves have a similar range in $\log p_k$, we measure the ratio $\theta$ of
the area between the curves and the area (in the $\log p_k,\log
k$-space) spanned by the extreme values
of $\log k$ and $\log p_k$. We find $\theta=0.95\%$ for our model, $4.0\%$ for
the BA model, and $11\%$ for the FKP model. Although both the BA and FKP
models hae been extended to yield better data fits~\cite{yook:inet,alv:inet}, 
the original forms of the models illustrate two important components of
Internet growth, namely 
the rich-gets-richer effect driving the growth of the BA model and
the spatial trade-off effect of the FKP model. 

A combination of these effects may explain why our model's degree
distribution, and the curve of the real network, lies between those of
the original BA and FKP models. In our model, the degrees of nodes do
not directly affect the creation of new links. However, 
preferential attachment occurs indirectly via 
positive feedback---nodes with large
degree acquire more traffic, and thus more budget which they can
reinvest in more connections, thus increasing their degree.
The effect of preferential attachment in the model is
shown in Fig.~\ref{fig:deg}(d), which is a plot of the probability
density of a node's traffic load given its degree.  Because an agent's
income is correlated with the traffic that it propagates, and a larger
budget will increase the possibility of creating new links, there is
positive feedback between the degree and the rate of degree increase,
i.e.\ a form of preferential attachment. Note that the
correlation in Fig.~\ref{fig:deg}(d) is not linear (the slope is
different from the solid line's). It is known that nonlinear
preferential attachment does not give a power-law degree
distribution~\cite{krap:prefatt} (which we seem to have), so
preferential attachment is not the only factor affecting our network's
growth. (If we had linear preferential attachment, the slope of $P(k)$
would, furthermore, be the same as the BA model.)

\begin{figure}
  \centering\resizebox*{0.95\linewidth}{!}{\includegraphics{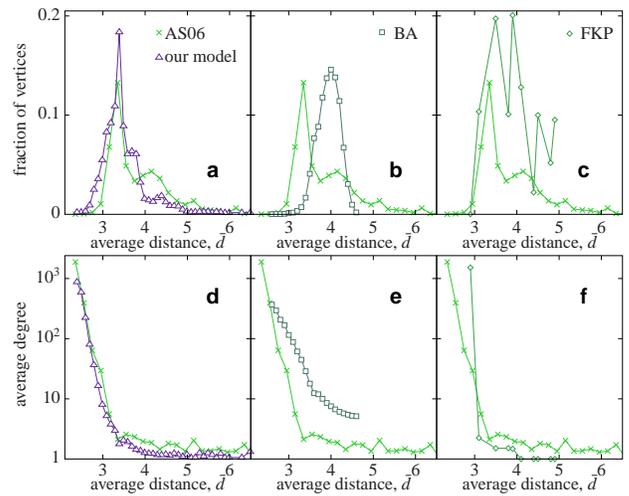}}
  \caption{Radial statistics for real and model networks. Panels (a)--(c)
    show the radial densities of nodes for the real AS-graph and our
    algorithm (a), the BA (b) and FKP (c) model. Panels (d)--(f) show the
    average degree vs. average distance $\bar{d}$ for our algorithm,
    the BA, and the FKP model respectively. The data of panels (b),
    (c), (e), and (f) are plotted in Ref.~\protect\cite{our:rad} as well. }
  \label{fig:rad}
\end{figure}

\subsection{Radial structure}

Structually, the AS graph is hierarchically
ordered~\cite{rex:hierarchy}---engineers and network operators speak
of the first, second and third tier.  For the model networks, we
measure a node's position in the hierarchy by its network
centrality~\cite{our:rad}. In Fig.~\ref{fig:rad} we diagram the
average fraction of nodes and the average degree as functions of the
average distance $\bar{d}$ to other nodes in the network ($\bar{d}$ is
the inverse of a centrality measure, known as \textit{closeness
  centrality}, so more central nodes are to the left in the
diagrams). By this method we can get a \textit{radial} picture of the
AS graph structure from the center to the periphery.  In
Fig.~\ref{fig:rad}(a)--(c) we plot the fraction of vertices at
different $\bar{d}$-values.  We note that our model resembles the real
AS-graph more closely than the BA and FKP models.  Having peaks
(roughly corresponding to the tiers of the Internet) like the observed
AS-graph. The shift to the left of the model curve in
Fig.~\ref{fig:rad}(a) can, to some extent, be explained by its smaller
size  (larger networks have larger average
distances, leading to a curve displaced to the right). In brief, the
BA model lacks the complex periphery of the real AS-graph (the density
is more balanced, compared with the left-skewed curve of the
real-world network). The average degree as a function of $\bar{d}$ is
less right-skewed in the BA model compared with the empirical
network. Just like the degree distribution, the FKP model deviates
from the real network in the opposite way compared to the BA
model---the high degree nodes of the FKP model are extremely
concentrated to the center of the network.

\begin{figure}
  \centering\resizebox*{.9 \linewidth}{!}{\includegraphics{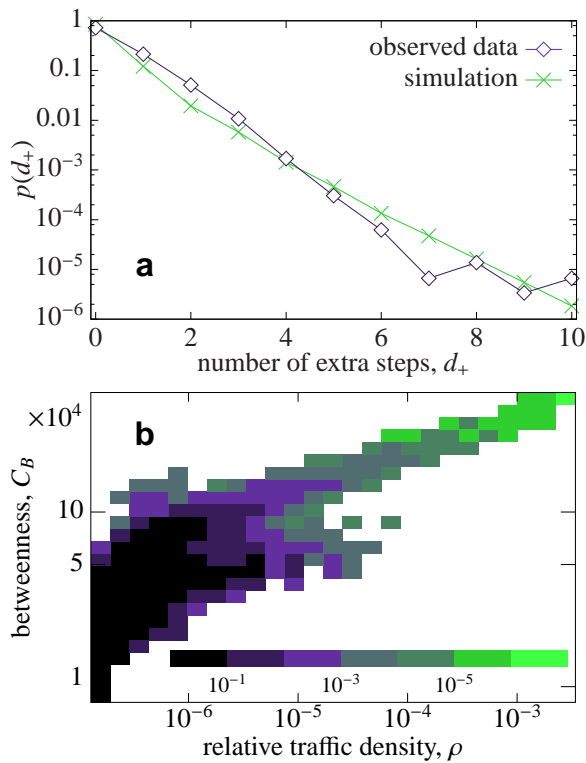}}
  \caption{Traffic patterns of the model. (a) displays the number of
    extra steps $d_+$ in packet navigation in the real Internet
    compared to our model. Panel (b) shows the probability density of
    agents having betweenness $C_B$ and traffic density $\rho$. The
    data is collected from twenty independent runs.}
  \label{fig:lbc}
\end{figure}

\subsection{Traffic flow and congestion patterns}

In Section~\ref{subsec:growth_results} we investigated network
topology and its growth. In this section we study traffic flow and how
network topology affects it.  In the Internet, packets do not
necessarily travel the shortest distances between source and
destination. Most importantly, business agreements between agents
arrange agents into a hierarchy~\cite{gao:relation}. The business
contracts put constraints on how packets are routed---for example,
usually a packet cannot first be routed downwards (to customers), then
upwards (to providers), in the hierarchy, even if that is a shorter
path (known as the valley free rule). Gao and Wang~\cite{gao:inflate}
investigated the extra distance $d_+$ packets need to travel due to
such reasons. They found a decaying probability distribution of $d_+$,
meaning that most of the traffic actually travels via shortest
paths. In our model we do not have explicit business agreements that
cause hierarchical routing into the core of the network, and out
again. It is, however, true for most graphs that a vast majority of
shortest paths pass a restricted core of the graph~\cite{goh:sfpnas},
and our traffic model routes most traffic via short (if not the
shortest) paths. The $d_+$ distribution of our model (shown in
Fig.~\ref{fig:lbc}(a)) matches the observation of Gao and
Wang~\cite{gao:inflate}.

We proceed to investigate the relationship between graph centrality
and traffic density. This can tell us something about how congestion
and fluctuations affect routing~\cite{holme:traffic}. If all
agents have sufficient capacity for packets to always route along
shortest paths, then traffic density along a link $l$ will be
proportional to its \textit{betweenness centrality}
\begin{equation}
C_B(l)=\sum_{i,j} \sigma_l(i,j)\; \Big / \sum_{i,j}\sigma(i,j)
\end{equation}
where $\sigma_l(i,j)$ is the number of shortest paths between nodes
$i$ and $j$ passing through the link $l$, and $\sigma(i,j)$ is the
total number of shortest paths between $i$ and $j$. If an AS is
congested, the traffic through its links will be lower than
anticipated by the betweenness of the edge. Thus, congestion patterns
can be illustrated by studying betweenness and traffic
load. Fig.~\ref{fig:lbc}(b) is a density plot of the actual traffic
density as a function of betweenness of the links of the model
network.  For more central nodes (higher betweenness), there is a
strong correlation between betweenness and traffic density---the
vertices with $C_B\approx 4\times 10^5$ spans half a decade of
$\rho$. For the more peripheral nodes the correlation is less clear
(vertices with $C_B\approx 5\times 10^4$ can have $\rho$-values of
almost three orders of magnitude). Indeed, there seems to be a
separation of agents into two classes, one with capacity to keep the
traffic flowing, another with too low capacity. For links of low
betweenness the traffic--betweenness correlation is weak. To
summarize, congestion does affect the system, and it is most
pronounced for nodes carrying little, or intermediate, traffic levels.

\begin{figure*}
  \centering\resizebox*{.5 \linewidth}{!}{\includegraphics{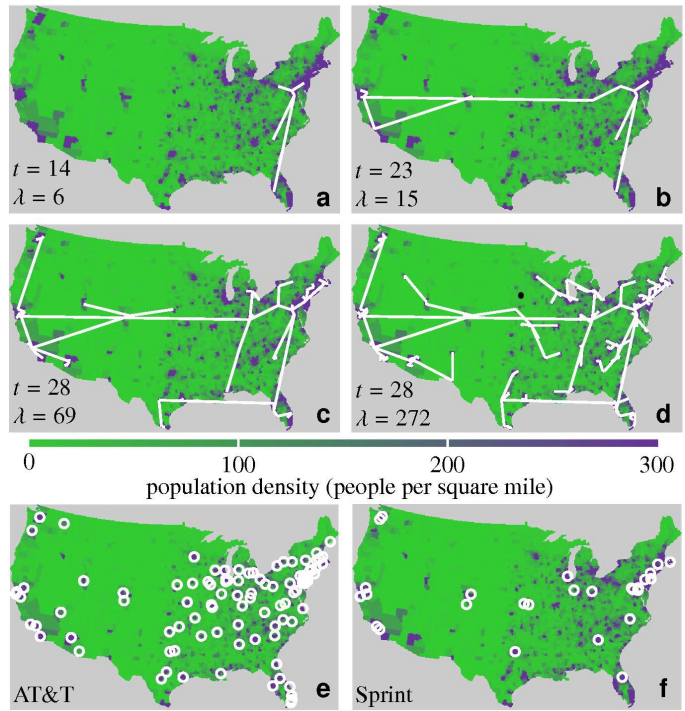}}
  \caption{The spatial expansion of a single agent with the US
    population density as model input. The simulation parameters are
    the same as the rest of the paper, except $n_D=20$, $L_x=513$ and
    $L_y= 323$. Panels (e) and (f) represent the points of presence of
    AT\&T and Sprint within the United States.  This data was adapted
    from Ref.~\protect\cite{rocket}.}
  \label{fig:usa}
\end{figure*}

\subsection{Geographic structure}

We briefly discuss the spatial network structure---another feature
that emerges from our model. As an example, we ran the simulation on
the population density profile of the United States. In
Fig.~\ref{fig:usa}(a)--(d) we show the growth of the largest agent for
a run with $n_D=20$, $L_x=513$ and $L_y= 323$. Lines are drawn between
each node (pixel) and the agent's nearest node at the time of the
node's addition. In this representation the length of the lines are
proportional to the wire cost.  Fig.~\ref{fig:usa}(e) and (f) plot the
locations of Tier~1 exchange points of two major Internet providers
Sprint and AT\&T (adapted from Ref.~\cite{rocket}). There are some
similarities between these real networks and the model network of
Fig.~\ref{fig:usa}(d)---all networks span the whole continent and have
locations concentrated in urban areas. In future work we intend
to make a statistical characterization of the spatial aspects of the
networks produced by our model.

\section{Discussion}

We have presented a model of communication networks that, like the
AS-level Internet, is built of spatially extended subnetworks that
have an interest in increasing the traffic running through them. Our
model networks grow slowly until they reach a critical mass where an
approximately exponential growth begins; they match the degree
distribution of real networks and the radial statistics
closely. The degree distribution of the model, and the real world lies
between the distributions of the pure BA and  FKP models. Since the model
incorporates aspects of both the BA and FKP models we hypothesize
that, the explanation for the degree distribution of the model, and
the real world, is a combined result of preferential attachment (of
the BA model) and geographically constrained optimization (of the FKP
model). We are able to recreate the traffic characteristic observed in
real Internet traffic. If we run the model on the US population
density map many features of the backbone of large, real agents are
recreated.

The different aspects of the model (traffic, geography, and agents
trying to increase the traffic they relay) all affect the output. In
this paper we do not scrutinize the model's parameter dependence, although
preliminary studies indicate that the speed of growth (quantified by
e.g.\ the time to reach the critical density) is strongly dependent on
both the wire and attachment prices, the population density profile (a
more clumped population distribution produces faster growth), and their
desire to communicate. On the other hand, the network topology is
rather insensitive to the population distribution, and also not very
dependent on how sources and destinations are generated (e.g.,
introducing a distance dependence does not matter much). The specific
layout of the network is, however, dependent on population
profile.

Many interesting extensions of the basic model are possible. One
interesting extension would, for example, be to include business
agreements between the different agents (similar to
Ref.~\cite{claffy:aseco,peer:chang}), or change the traffic patterns
from the person--to--person communication of the present model to a
situation with more traffic coming from central servers.  It might
also be interesting to model intra-AS routing.  Many of today's ASes
employ ``hot-potato'' routing and transfer packets to the next AS as
quickly as possible, to reduce cost.  Alternative intra-AS routing
strategies, such as routing the packet as close to the destination as
possible, could be tested within the model's framework.

\subsection*{Acknowledgements}
The authors would like to thank Allen Downey for his helpful comments.
PH acknowledges financial support from the Swedish Foundation for
Strategic Research.  SF acknowledges the support of the National
Science Foundation (grants CCF 0621900 and CCR--0331580), and the
Santa Fe Institute.


\begin{thebibliography}{10}

\bibitem{alv:inet}
J.~I. Alvarez-Hamelin and N.~Schabanel.
\newblock An {I}nternet graph model based on trade-off optimization.
\newblock {\em Eur. Phys. J. B}, 38:231--237, 2004.

\bibitem{bar:inet}
S.~Bar, M.~Gonena, and A.~Wool.
\newblock A geographic directed preferential {I}nternet topology model.
\newblock {\em Computer Networks}, 51:4174--4188, 2007.

\bibitem{ba:model}
A.-L. Barab\'{a}si and R.~Albert.
\newblock Emergence of scaling in random networks.
\newblock {\em Science}, 286:509--512, 1999.

\bibitem{Bonabeau02a}
E.~Bonabeau.
\newblock Agent-based modeling: Methods and techniques for simulating human
  systems.
\newblock {\em Proc Natl Acad Sci}, 99:7280--7287, 2002.

\bibitem{cairncross}
F.~Cairncross.
\newblock {\em The death of distance}.
\newblock Harvard Business School Press, Boston, MA, 1997.

\bibitem{carlson:hot}
J.~M. Carlson and J.~Doyle.
\newblock Highly optimized tolerance: a mechanism for power laws in designed
  systems.
\newblock {\em Phys. Rev. E}, 60:1412--1427, August 1999.

\bibitem{chang:superhot}
H.~Chang, S.~Jamin, and W.~Willinger.
\newblock Internet connectivity at the {AS}-level: an optimization-driven
  modeling approach.
\newblock In {\em MoMeTools '03: Proceedings of the ACM SIGCOMM workshop on
  Models, methods and tools for reproducible network research}, pages 33--46,
  New York, NY, USA, 2003. ACM.

\bibitem{peer:chang}
H.~Chang, S.~Jamin, and W.~Willinger.
\newblock To peer or not to peer: Modeling the evolution of the {Internet's
  AS-level} topology.
\newblock In {\em Proc. IEEE INFOCOM}, 2006.

\bibitem{aaron:power_law}
A.~Clauset, C.~R. Shalizi, and M.~E.~J. Newman.
\newblock Power-law distributions in empirical data.
\newblock e-print arXiv:0706.1062, 2007.

\bibitem{chn:perc}
R.~Cohen, K.~Erez, D.~ben Avraham, and S.~Havlin.
\newblock Resilience of the {I}nternet to random breakdowns.
\newblock {\em Phys. Rev. Lett.}, 85:4626--4628, 2000.

\bibitem{daub:as}
I.~Daubechies, K.~Drakakis, and T.~Khovanova.
\newblock A detailed study of the attachment strategies of new autonomous
  systems in the {AS} connectivity graph.
\newblock {\em Internet Mathematics}, 2:185--246, 2006.

\bibitem{echen:cong}
P.~Echenique, J.~{G\'{o}mez-Gard\~{e}nes}, and Y.~Moreno.
\newblock Dynamics of jamming transitions in complex networks.
\newblock {\em Europhys. Lett.}, 71:325--331, 2005.

\bibitem{fkp:model}
A.~Fabrikant, E.~Koutsoupias, and C.~H. Papadimitriou.
\newblock Heuristically optimized trade-offs: A new paradigm for power laws in
  the {I}nternet.
\newblock In {\em Proceedings of the 29th International Conference on Automata,
  Languages, and Programming}, volume 2380 of {\em Lecture notes in Computer
  science}, pages 110--122, Heidelberg, 2002. Springer.

\bibitem{f3}
M.~Faloutsos, P.~Faloutsos, and C.~Faloutsos.
\newblock On power-law relationships of the {I}nternet topology.
\newblock {\em Comput. Commun. Rev.}, 29:251--262, 1999.

\bibitem{gao:relation}
L.~Gao.
\newblock On inferring autonomous system relationships in the {I}nternet.
\newblock {\em IEEE / ACM Transactions on Networking}, 9:733--745, 2001.

\bibitem{gao:inflate}
L.~Gao and F.~Wang.
\newblock The extent of {AS} path inflation by routing policies.
\newblock In {\em Proceedings of GLOBECOM '02}, volume~3, pages 2180--2184,
  2002.

\bibitem{goh:sfpnas}
K.-I. Goh, E.~Oh, H.~Jeong, B.~Kahng, and D.~Kim.
\newblock Classification of scale-free networks.
\newblock {\em Proc. Natl. Acad. Sci. USA}, 99:12583--12588, 2002.

\bibitem{holme:traffic}
P.~Holme.
\newblock Congestion and centrality in traffic flow on complex networks.
\newblock {\em Advances in Complex Systems}, 6:163--176, 2003.

\bibitem{our:rad}
P.~Holme, J.~Karlin, and S.~Forrest.
\newblock Radial structure of the {I}nternet.
\newblock {\em Proc. R. Soc. A}, 463:1231--1246, 2007.

\bibitem{isard}
W.~Isard.
\newblock {\em Location and space economy}.
\newblock MIT Press, Cambridge MA, 1956.

\bibitem{krap:prefatt}
P.~L. Krapivsky, S.~Redner, and F.~Leyvraz.
\newblock Connectivity of growing random networks.
\newblock {\em Phys. Rev. Lett.}, 85:4629 -- 4632, 2000.

\bibitem{vesp:inet}
R.~Pastor-Santorras and A.~Vespignani.
\newblock {\em Evolution and structure of the {I}nternet: a statistical physics
  approach}.
\newblock Cambridge Univeristy Press, Cambridge, 2004.

\bibitem{rfc1771}
Y.~Rekhter and T.~Li.
\newblock {A Border Gateway Protocol 4 (BGP--4)}.
\newblock RFC 1771 (Draft Standard), Mar. 1995.
\newblock Obsoleted by RFC 4271.

\bibitem{claffy:aseco}
S.~Shakkottai, T.~Vest, D.~Krioukov, and K.~C. Claffy.
\newblock Economic evolution of the {Internet} {AS}-level ecosystem.
\newblock e-print arxiv:cs.NI/0608058, 2006.

\bibitem{sood:rw}
V.~Sood and P.~Grassberger.
\newblock Localization transition of biased random walks on random networks.
\newblock {\em Phys. Rev. Lett.}, 99:098701, 2007.

\bibitem{rocket}
N.~Spring, R.~Mahajan, D.~Wetherall, and T.~Anderson.
\newblock Measuring {ISP} topologies with {R}ocketfuel.
\newblock {\em IEEE / ACM Transactions of Networking}, 12:2--16, 2004.

\bibitem{rex:hierarchy}
L.~Subramanian, S.~Agarwal, J.~Rexford, and R.~H. Katz.
\newblock Characterizing the {I}nternet hierarchy from multiple vantage points.
\newblock In {\em INFOCOM 2002. Twenty-First Annual Joint Conference of the
  IEEE Computer and Communications Societies. Proceedings. IEEE}, volume~2,
  pages 618--627, 2002.

\bibitem{yook:inet}
S.-H. Yook, H.~Jeong, and A.-L. Barab\'{a}si.
\newblock Modeling the {I}nternet's large-scale topology.
\newblock {\em Proc. Natl. Acad. Sci. USA}, 99:13382--13386, 2002.

\end{thebibliography}
\end{document}